\documentclass{emulateapj}

\usepackage{amssymb}

\slugcomment{ApJSup accepted (Spitzer Special Issue)}
\shorttitle{Infrared Imaging of High Redshift Submillimeter Sources}
\shortauthors{Charmandaris et al.}

\begin{document}


\title{Imaging of High Redshift Submillimeter Galaxies at 16 and 22\,$\mu m$
with the Spitzer/IRS\footnote{The IRS was a collaborative venture
between Cornell University and Ball Aerospace Corporation funded by
NASA through the Jet Propulsion Laboratory and the Ames Research
Center.}~Peak-up Cameras: Revealing a population at
z$>$2.5\footnote{Based on observations obtained with the Spitzer Space
Telescope, which is operated by JPL, California Institute of
Technology for the National Aeronautics and Space Administration.}}

\author{V. Charmandaris\altaffilmark{3,4}, K.I. Uchida\altaffilmark{3},
D. Weedman\altaffilmark{3}, T. Herter\altaffilmark{3},
J.R. Houck\altaffilmark{3}, H.I. Teplitz\altaffilmark{5},
L. Armus\altaffilmark{5}, B.R. Brandl\altaffilmark{6},
S.J.U. Higdon\altaffilmark{3}, B.T. Soifer\altaffilmark{5},
P.N. Appleton\altaffilmark{5}, J. van Cleve\altaffilmark{7},
J.L. Higdon\altaffilmark{3}}

\altaffiltext{3}{Astronomy Department, Cornell University, Ithaca, NY 14853}

\altaffiltext{4}{Chercheur Associ\'e, Observatoire de Paris, F-75014, Paris, France}
\altaffiltext{5}{Spitzer Science Center, California Institute
of Technology, 220-6, Pasadena, CA 91125}

\altaffiltext{6}{Leiden University, 2300 RA Leiden, The Netherlands}

\altaffiltext{7}{Ball Aerospace and Technologies Corp. 1600 Commerce St., Boulder, CO 80301}

\email{vassilis@astro.cornell.edu}

%\newpage

\begin{abstract}
  
  We present broad band imaging observations obtained with the ``peak
  up'' imagers of the Spitzer Space Telescope Infrared Spectrograph
  (IRS) at wavelengths of 16\,$\mu$m and 22\,$\mu$m for a number of
  sources detected primarily at submillimeter wavelengths, which are
  believed to be at high, though undetermined, redshift. We targeted
  11 sources originally detected by SCUBA and 5 submillimeter sources
  detected at 1.2mm by MAMBO.  Two optically discovered quasars with
  z$>$6 were also observed to determine if there is detectable dust
  emission at such high redshifts.  Seven of the submillimeter sources
  and both high-redshift quasars were detected, and upper limits of
  about $\sim$50 $\mu$Jy apply to the remainder.  Using their
  mid-/far-IR colors, we demonstrate that all of the submillimeter
  sources are at z$>$1.4. The mid-IR colors for two of our detections
  and three of our strong upper limits suggest that these galaxies
  must be at z$>$2.5.  We also introduce a technique for estimating
  redshifts and source characteristics based only on the ratio of
  fluxes in the 16\,$\mu$m and 22\,$\mu$m images.

\end{abstract}

\keywords{dust, extinction ---
  infrared: galaxies ---
  galaxies: submillimeter ---
  galaxies: high-redshift --
  galaxies: AGN ---
  galaxies: starburst}

\section{Introduction}

It is currently widely accepted that the energy production in the
universe, often quantified with the star formation rate (SFR),
increases by more than an order of magnitude as we move from the local
Universe to z$\ge$2 \citep[e.g.,][]{Madau98}. Interstellar dust, a
direct product of star formation and evolution, often obscures and
absorbs the light emitted from stars or non-thermal sources and
reradiates at longer wavelengths in the mid- and far-infrared
(mid-IR/far-IR) and submillimeter. The results of IRAS, launched more
than 20 years ago, have demonstrated that in the local universe 6\% of
the total infrared emission originates from the centers of galaxies
which are highly enshrouded by dust and therefore are extremely
difficult to probe through direct observations in the optical and
near-infrared (near-IR). These are the luminous and ultraluminous
infrared galaxies (LIRGs/ULIRGs) \citep[e.g.,][]{Houck84}. Deep IR
extragalactic surveys, mainly performed at 15$\mu$m with the Infrared
Space Observatory (ISO), recently revealed a strong evolution in
galaxy counts such that at z$\sim$1 a large fraction of the cosmic
infrared background (CIRB) is due to LIRGs and ULIRGs
\citep{Franceschini01,Elbaz02a,Elbaz03,Metcalfe03}. Furthermore, deep
maps with the Sub-mm Common User Bolometric Array (SCUBA), as well as
studies with the Max-Plank Millimeter Bolometer (MAMBO), have shown
that almost 100\% of the submillimeter background can be resolved into
individual sources, most of which are too faint for optical
spectroscopy \citep[see][ for a review]{Blain02}.

Unlike the majority of ISO sources detected at 15$\mu$m, for which
direct optical imaging and redshift measurements were feasible
\citep[see][]{Franceschini01,Elbaz02a}, the nature of the
submillimeter galaxy (SMG) population is much more elusive. More than
300 SMGs have been detected to date, but their faintness in the
optical and near-IR and the positional uncertainty of the $\sim$13$''$
SCUBA beam made a secure identification of a counterpart at those
wavelengths very challenging.  The typical spectral energy
distribution (SED) of a dusty galaxy produces a negative K correction
in the mm and sub-mm for high redshift systems, making detection of
such sources much easier than at shorter wavelengths.  The observed
surface density of SCUBA sources and estimated photometric redshifts
imply that the co-moving density of a high redshift,
$\sim10^{13}$\,L$_{\sun}$ population of galaxies is $\sim$400 times
greater than in the local universe \citep{Blain02}.  Confirmation that
such high-redshift sources are present was achieved by
\citet{Chapman03} who used optical spectroscopy and accurate radio
continuum positions \citep{Barger00, Ivison02} to measure the
redshifts for ten such sources and show that for this population the
median redshift is $<z>\sim$2.4.

It remains crucial to develop diagnostics for understanding the nature
of these luminous, primordial sources and to determine accurate
redshifts so that their evolution can be traced. Optical observations
probe the rest-frame ultraviolet or blue continuum, which only
characterizes the small fraction of luminosity that escapes the dust.
Observations at longer rest frame wavelengths can place important
constrains on the properties of dust as well as the physics of the
dominant ionizing source and are essential if the optical is invisible
\citep[i.e.][]{Frayer04}.  Near-IR spectroscopy even with 8m telescopes
is extremely difficult for SMGs, which typically have an m$_{\rm
  I}>22$mag, so the mid-IR is the next available wavelength window.
Combining the results of the deep surveys performed with ISO and SCUBA
over the past 7 years, only 11 galaxies have been observed at both 15$\mu$m
and 850$\mu$m, and redshifts have been determined for 9 of them (see
Fig.1a). Seven of the SMGs were detected because of favorable
gravitational amplification of their emission by foreground lensing
clusters \citep{Metcalfe03}, two were identified in the CFRS14 field
\citep{Lilly99,Flores99, Higdon04}, one was found in the Hubble Deep
Field \citep{Hughes98, Aussel99}, and one, HR10, is an extremely red
object (ERO) at z=1.44 \citep{Elbaz02b}.

The superb sensitivity of the Spitzer Infrared Spectrograph (IRS)
\citep{Houck04}, and the ability of the IRS to efficiently obtain 16$\mu$m
and 22$\mu$m images to levels of several tens of $\mu$Jy and follow-up
spectra to a level of $\sim$0.5mJy opens a unique opportunity to
examine the mid-IR emission from SMGs. In this paper we present our
first results from a program of targeted observations of a number of
such sources. We discuss selection criteria and observing strategy in
section 2, results are presented in section 3, and our conclusions can
be found in section 4.

\section{Observations}

The targets were selected from a number of recent submillimeter
catalogs, with several objectives.  Primarily, we wished to obtain
mid-IR images of sources for which no positive identification had been
made with an optical counterpart, or for which no redshift had been
obtained but was expected to be z$>$1. Various sources with no optical
counterparts will be found by Spitzer deep surveys in the mid-IR, so
it is important to begin calibrating the properties of samples derived
in other ways which may have similarities to such sources. The
16$\mu$m to 850$\mu$m ratios can improve estimates of the probable
redshift. We can also utilize the 16$\mu$m to 22$\mu$m flux ratio to
determine the slope of the infrared spectrum or the presence in one or
the other band of any strong polycyclic aromatic hydrocarbon (PAH)
emission features or silicate absorption features.  The final
objective of our imaging survey of SMGs was to determine whether any
are bright enough for an IRS spectroscopic follow-up which can give a
precise redshift measurement\footnote{For f$_{22\mu m}>$0.7mJy we can
  obtain a 20--38$\mu$m spectrum with a $>5\sigma$ detection in
  $\sim$1hr.}.  In addition to observing the submillimeter sources, we
added two quasars at very high redshift (z $>$ 6) identified by the
Sloan Digital Sky Survey to determine if such distant quasars also
have sufficient re-emission from dust to be detectable in Spitzer deep
surveys.  One of these quasars, J1148+5251, has many characteristics
of the extreme submillimeter sources; it has
$\sim2\times10^{10}$\,M$_{\sun}$ of molecular gas and an
L(FIR)$\sim1.3\times10^{13}$\,L$_{\sun}$ \citep{Walter03}, has been
detected by MAMBO (f$_{1.2mm}$=5.0$\pm$0.6mJy), and is estimated to
contain $\sim7\times10^{8}$\,M$_{\sun}$ of dust \citep{Bertoldi03}.
Whether sufficiently hot dust is present in such AGN to be seen at the
much shorter wavelengths observed in the rest frame is important to
know for interpreting other sources.  All of our targets were observed
during the first 3 IRS campaigns that took place between 14 December
2003 and 9 February 2004.

We used the blue and red peak up cameras located in the short low
module of the IRS to observe our targets at 16$\mu$m
(13.5--18.7$\mu$m) and 22 $\mu$m(18.5--26.0$\mu$m).  By using the
Spitzer focal plane as the reference frame, it is possible to program
observations to obtain dithered observations in both the 16$\mu$m and
22 $\mu$m images as a standard IRS short low first order (SL1) staring
AOT \citep[see][]{Houck04} with the targets defined of type ``fixed
cluster offsets'' on the plane of the array.  The offsets, in arcsecs
parallel and perpendicular to the SL1 slit, were set to (-62$''$,
289$''$) and (50$''$, 284$''$) for the centers of the blue and red
peak up windows respectively.  When the telescope was performing the
standard nodding pattern on the SL1 slit, the actual target was imaged
at two positions on the peak up cameras separated by 18$''$.  This
approach enables us to obtain both images with a single telescope
pointing, and also to collect the data in the ``raw mode'', which
allows flexible image analysis.

The data were processed by the standard IRS pipeline (version S9.1) at
the Spitzer Science Center (see chapter 7 of Spitzer Observing
Manual\footnote{http://ssc.spitzer.caltech.edu/documents/som/}).  The
2D images were converted from slopes after linearization correction,
subtraction of darks, and cosmic ray removal. The resulting images
were divided by the photometric flat, and a world coordinate system
was inserted into them using the reconstructed pointing of the
telescope. The astrometric accuracy of our images is better than
$\sim$2$''$ and the FWHM of the point spread function is 3.5$''$ and
4.8$''$ at 16$\mu$m and 22$\mu$m respectively. Since we typically had
multiple images of the same target, we used the astrometry information
to shift and co-add our pointings, as well as to perform sky
subtraction by taking the differences of the two nod positions. Flux
calibration was performed using a number of stars for which high
quality spectral templates were available. The properties of all
targets observed are presented in Table 1.  Our photometric
uncertainties are less than $\sim$10\%. The detection limit of the
peak up cameras depends rather strongly on the brightness of the
background. Since the targets observed are in a low background region
with surface brightness $\sim$20MJy\,sr$^{-1}$, the 1$\sigma$
detection limit is $\sim$30$\mu$Jy for an integration time of 120sec
at 16$\mu$m and 240sec at 22$\mu$m.

\section{Results}

As we show in Table 1, we were able to detect 10 out of the 16
submillimeter targets at $>3\sigma$, including all 3 sources for which
there are published redshifts. Only one source, CFRS 10.1411 at just
z=0.07 \citep{Lilly99}, is above 1mJy. The fluxes of the remainder of
the SMG detections are on the order of a few hundred $\mu$Jy in both
16 and 22$\mu$m. No source without a known redshift is above the
f$_{22\mu m}$=0.7mJy.  The two quasars, J1048+4637 and J1148+5251,
with redshifts of 6.23 and 6.42, respectively, were also detected.
Given their extreme redshift, our measurements probe their rest frame
near-IR emission ($\sim$2--3$\mu$m) which is likely due to a hot dust
component.  This is especially significant for J1148+5251, a strong
submillimeter source, because it could be used to explore whether AGN
at high redshift which have sufficient cool dust for submillimeter
emission can contain a hot dust component.  It also illustrates the
challenges in determining the primary luminosity source (starburst or
AGN) for the ``submillimeter'' population.

When we combine our measurements from the IRS peak up filters with the
SCUBA 850$\mu$m points, we can bracket the peak of the infrared SED of
a source, typically found at $\sim$80$\mu$m in the rest frame, and
obtain a photometric estimate of its redshift. In a manner similar to
the discussion of \cite{Lilly99} in exploring the ISO 15$\mu$m to
850$\mu$m colors, we calculated the f$_{16 \mu m}$/f$_{850 \mu m}$
ratio as a function of z for a number of various template SEDs. Our
results for a few selected templates are presented in Figure 1a. We
can see that as the source moves to higher z, due to the negative K
correction the flux sampled by SCUBA at 850$\mu$m remains relatively
flat or decreases only slightly. The rate of the decrease clearly
depends on where the far-IR SED peaks and it is faster(slower) when
the galaxy has warmer(colder) far-IR color. The flux measured by our
16$\mu$m peak up window decreases much more rapidly up until z$\sim$1
when the rest frame $\sim$6--8$\mu$m range begins to enter the window.
Beyond that point the details depend on features in the mid-IR
spectrum of the source. If a source has a featureless mid-IR continuum
with weak PAH emission such as an AGN dominated template (i.e.
NGC\,1068), the f$_{16 \mu m}$ flux would decrease monotonically.  The
situation is a bit more complex if strong PAH emission features (such
as those seen in M82) or silicate absorption (as in the case of
Arp\,220) are present in the spectrum.  It is evident from Figure 1a
that given a sufficiently large sample of possible templates we can
define a lower limit to the possible f$_{16 \mu m}$/f$_{850 \mu m}$ at
a given redshift. Calculating this ratio as a function of z for a wide
range of known SEDs suggests that at any given z$>$0.5, where most
SMGs are found, an SED similar to that of Arp\,220 provides the most
conservative limit of the possible value f$_{16 \mu m}$/f$_{850 \mu
  m}$.  By measuring this ratio for our SMG targets, we can obtain the
absolute {\em lower limit} for their redshift, assuming this SED.

To demonstrate this, we also plot in Figure 1a, along with our data,
all SMG sources for which redshifts and ISO 15$\mu$m fluxes have been
published to date\footnote{Given the similarities between the ISO
  15$\mu$m (LW3) and the IRS 16$\mu$m blue peak up filters, as well as
  the possible shapes of the mid-IR SEDs for the ISO detected SMGs we
  have set their f$_{16 \mu m}$ $\sim$f$_{15 \mu m}$ $\pm$10\%.}. We
see that all SMGs of known redshift have f$_{16 \mu m}$/f$_{850 \mu
  m}$ ratios placing them between Arp\,220 and Mrk\,273.  An accurate
estimate of the redshift clearly depends on the assumed SED of the
source.  Our conservative method indicates that with one
exception\footnote{We cannot explain at this point why source \#9
  despite its strong detection and the fact that radio/sub-mm data
  suggest a z$\sim$2.4 \citep{Barger00}, has rather peculiar mid-IR
  and sub-mm colors.} all SMGs in Table 1 with no known redshift must
be at z$>$1.4.  Even if at z$\sim$1.5, a source like Arp\,220 would
have an m$_{\rm I}\sim$23.5mag, well within spectroscopic capabilities
from the ground. Our upper limits for the f$_{16 \mu m}$/f$_{850 \mu
  m}$ ratio on sources \#4, \#10 and \#14 suggest that these sources
are beyond z$>$2.5.

Interestingly, the highest redshift (z=2.8) known ISO source, SMM
02399-0136 in Abell 370 \citep{Metcalfe03}, as well as our target \#1,
the ERO N4 of \citet{Frayer03}, are located very close to the pure AGN
locus (NGC\,1068). If the SMGs are indeed warm AGN-like sources
following the SEDs discussed by \citet{Blain99}, then our Arp\,220
limit is extremely conservative.  In this case, the luminosity of
these sources would derive from a scaled up version of an enshrouded
AGN, and the thermal emission from the AGN in the 1--6$\mu$m rest
frame will enter into our 13$\mu$m--18$\mu$m blue peak up window at z$>$2.
This will increase the expected f$_{16 \mu m}$/f$_{850
  \mu m}$ ratio by a factor of $\sim$8, as seen in Figure 1a, so that
the necessary redshifts will easily exceed z$\sim3$.

We can also begin to investigate whether it will eventually prove
possible to get reliable photometric redshifts from the 16$\mu$m and
22$\mu$m imaging data alone, and to use this mid-infrared flux ratio
for diagnostic information on the dominance of an AGN or a starburst.
The concept of this is illustrated in Figure 1b.  Galaxies whose
mid-IR luminosities derive primarily from starbursts are characterized
by strong PAH emission features \citep[e.g.][]{Brandl04}, whereas AGN
generally produce dust continua without PAH emission but often with
silicate absorption \citep[e.g.][]{Laurent00}.  While we intend to
calibrate the effect with many more spectroscopic observations of
known starbursts and AGN, we illustrate in Figure 1b the changes that
would be expected in the f$_{16 \mu m}$/f$_{22 \mu m}$ ratio for
templates like the ISO spectra of the starburst M82 \citep{Sturm00},
the AGN NGC\,1068 \citep{LeFloch01}, and the more enshrouded sources
Arp\,220 \citep{Spoon04} and Mrk\,273 \citep{Devriendt99}.  Note that
even though NGC\,1068 hosts a Sy2 nucleus affected by considerable
extinction in the optical, it displays a nearly power-law continuum
with very weak PAH and silicate features in the mid-IR
\citep{Sturm00,LeFloch01}. Over a broad range of redshifts for z$>$1,
the approximately power-law spectrum of this AGN means, therefore,
that the 16$\mu$m to 22$\mu$m flux ratio has a narrow range, from 0.7
to 0.9, whereas this ratio for other sources with stronger spectral
features ranges between extremes of 0.1 to 1.7.  It is clear that this
flux ratio information is potentially useful for constraining source
characteristics and redshifts.  For illustration, we apply this
concept to the present results in Table 1 by plotting sources in
Figure 1b at the redshifts corresponding to the minimum redshift
determined from the f$_{16 \mu m}$/f$_{850 \mu m}$ ratio. Only sources
\#6, \#9, and \#11 have ratios considerably outside the expected
NGC\,1068 values.  Having the ratios shown implies that these 3
sources have strong spectral features and, for sources \#6 and \#11,
yields redshift estimates consistent with those derived from the
f$_{16 \mu m}$/f$_{850 \mu m}$ ratio in Figure 1a.

\section{Conclusions}

The unprecedented sensitivity of the IRS peak up cameras has enabled
us to perform efficient imaging of 16 SMGs at 16$\mu$m and 22$\mu$m.
Ten of these sources are detected, which doubles the number of mid-IR
detections for sources discovered through submillimeter surveys. Using
a number of SED templates along with the observed f$_{16 \mu
  m}$/f$_{850 \mu m}$ ratios or upper limits, we have shown that 13 of
these SMG sources must be located at z$>$1.4, with some being
constrained to z$>$2.5. We also introduce use of the observed f$_{16
  \mu m}$/f$_{22\mu m}$ ratio to constrain photometric redshifts, and
to demonstrate that some of the SMGs must have mid-infrared spectra
similar to those of AGN.

\acknowledgments 

VC would like to acknowledge H. Aussel and D. Elbaz (CEA, France) and
an anonymous referee for useful comments which improved the paper.
This work is based [in part] on observations made with the Spitzer
Space Telescope, which is operated by the Jet Propulsion Laboratory,
California Institute of Technology, under NASA contract 1407. Support
for this work was provided by NASA through Contract Number 1257184
issued by JPL/Caltech.

\begin{deluxetable}{ccccccrclll}
%\rotate
\tablewidth{0pc}
\footnotesize
\tabletypesize{\scriptsize}
\setlength{\tabcolsep}{0.0in}
\tablecaption{Mid-IR Properties of high-z sub-mm Galaxies\label{tbl1}}

\tablehead{
% first line of header
\colhead{RA}  & \colhead{DEC}  & \colhead{$f_{16 \mu m}$} &
\colhead{$f_{22 \mu m}$}  &  \colhead{$\frac{f_{16 \mu m}}{f_{22 \mu m}}$} & \colhead{z} & 
\colhead{$f_{850 \mu m}$} &  \colhead{$\frac{f_{16 \mu m}}{f_{850 \mu m}}$} &\colhead{[ID]/Comments} & \\
% second line of headeer
\colhead{(J2000)} & \colhead{(J2000)} & \colhead{(mJy)} & \colhead{(mJy)} & &  &\colhead{(mJy)} &  & & }
\startdata

04:43:07.1   & 02:10:25.1  & 0.52     & 0.98     & 0.54  &  2.51     &  7.2  &  0.072  & [1] SMM J04431+0210(N4) in (1)\\
10:00:37.2   & 25:14:57.9  & 1.88     & 2.55     & 0.73  &  0.07     &  4.7  &  0.400  & [2]\tablenotemark{a} CFRS 10.1411 in (2)\\ 
10:00:38.2   & 25:14:49.2  & 0.27     & 0.62     & 0.42  &  0.50     &  4.8  &  0.056  & [3] CFRS 10.1153 in (2)\\ 
12:36:06.4   & 62:11:38.0  & $<$0.026 & $<$0.030 &  -    & [$>$2.5] & 15.4  &  $<$0.002 & [4] HDFSMM-3606+1138 in (4) \\ 
12:36:18.4   & 62:15:49.2  & 0.12     & 0.13     & 0.92  & [$>$1.5] &  7.8  &  0.015  & [5] \#7 in (3) \\
12:36:46.7   & 62:14:46.9  & 0.13     & 0.38     & 0.34  & [$>$1.5] & 10.7  &  0.012  & [6] \#33 in (3)  \\
12:36:52.32  & 62:12:26.3  & $<$0.025 & $<$0.041 &   -   & [$>$2.0] &  7.0  &  $<$0.004  & [7] HDF N(850.1) in (5)\\ 
12:37:00.4   & 62:09:08.5  & 0.07     & 0.13     & 0.54  & [$>$1.8] & 11.9  &  0.006  & [8] \#49 in (3) \\  
12:37:09.5   & 62:08:39.5  & 0.70     & 0.61     & 1.15  & [$>$0.5] &  8.3  &  0.084  & [9] \#58 in (3)\\
12:37:29.7   & 62:10:51.0  & $<$0.028 & $<$0.024 &  -    & [$>$2.5] & 14.3  &  $<$0.002  & [10] HDFSMM-3730+1051 in (4)  \\ 
14:17:40.2   & 52:29:06.5  & 0.19     & 0.14     & 1.35  & [$>$1.4] &  8.8  &  0.022  & [11] CFRS 14A  in (6)\\ 
\hline
12:05:17.59  & -07:43:11.5 & $<$0.060 & $<$0.068 &  -    & [$>$1.6] &  6.3 & $<$0.009 & [12]\tablenotemark{a} MMJ 120517-0743.1 (NTT-MM25) in (7)\\
12:05:39.45  & -07:45:26.1 & $<$0.049 & $<$0.050 &  -    & [$>$1.7] &  6.3 & $<$0.008 & [13]\tablenotemark{a} MMJ 120539-0745.4 (NTT-MM16) in (7)\\
12:05:46.55  & -07:41:33.2 & $<$0.040 & $<$0.060 &  -    & [$>$2.2] & 18.5 & $<$0.003 & [14]\tablenotemark{a} MMJ 120546-0741.5 (NTT-MM31) in (7)\\
15:41:25.2   &  66:14:39.6 & 0.11     &  0.16    & 0.69  & [$>$1.6]   & 10.7 &  0.010  & [15] J 1541+6615 (A2125-MM26) in (7)\\
15:41:27.3   &  66:16:59.4 & 0.05     &  0.07    & 0.71  & [$>$2.2]   & 14.6 &  0.003  & [16] J 1541+6616 (A2125-MM27) in (7) \\
\hline
10:48:44.9   &    46:37:15.5    & 0.23  &   0.37   &    0.62   & 6.23   &   &   & [17] J1048+4637 \\
11:48:16.5   &    52:51:49.1    & 0.51  &   0.74   &    0.69   & 6.42   &   &   & [18] J1148+5251
\enddata

\tablenotetext{a}{Unlike the rest of the sources where the on source
integration times were 120sec and 240sec at 16$\mu$m and 22$\mu$m, the
corresponding times for these targets were 140sec and 280sec respectively.}

\tablecomments{In cases of non detections we indicate our 1$\sigma$
  upper limits. The identifications and more detail on individual
  sources are presented in: (1) \citet{Frayer03}, (2) \citet{Lilly99},
  (3) \citet{Barger00}, (4) \citet{Borys02}, (5) \citet{Hughes98}, (6)
  \citet{Eales99}, (7) \citet{Eales03}. Redshifts in brackets are
  absolute minimum redshifts based on photo-z and an Arp\,220 template
  (see section 3 and Fig.1).}

\end{deluxetable}
 
\begin{figure*}
\figurenum{1}
\epsscale{1.1}
\plotone{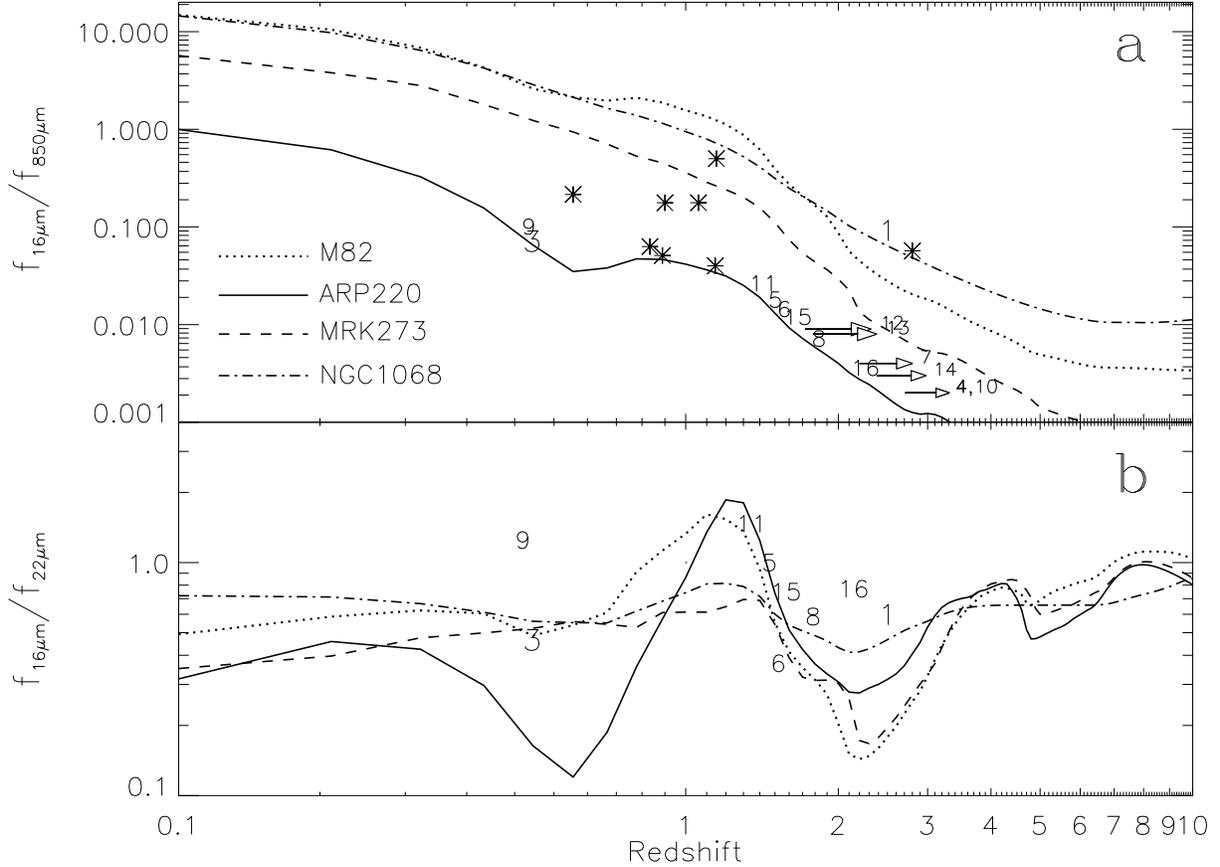}
 
\caption{ 
  a) The expected ratio of the flux densities f$_{16 \mu m}$/f$_{850
    \mu m}$ of a distant source as a function of redshift if the SED
  of the source is similar to a starburst like M82, an AGN like NGC
  1068, or an enshrouded source such as Arp\,220 or Mrk\,273.  Note
  that the SED of Arp\,220 defines in the most conservative way the
  {\em minimum redshift} of the SMG source for a given value of the
  ratio.  The minimum redshift for each SMG of our sample is indicated
  by the source number from Table 1. All previously known ISO 15$\mu$m
  sources at z$>$1 for which both redshifts and 850$\mu$m fluxes are
  available are plotted as asterisks.  Interestingly, the most distant
  ISO source as well as our source 1 for which the redshift is known,
  fall most closely to the AGN template.  Six sources (4,7,10,12,13,
  and 14) for which we only have upper limits at 16 and 22$\mu$m are
  indicated with arrows and must be at even higher redshifts (sources
  4,10, and 14 must be at z$>$2.5).  b) The expected flux ratio f$_{16
    \mu m}$/f$_{22 \mu m}$ as would be measured with the IRS peak up
  filters for the templates used above.  Note that if a galaxy has a
  starburst SED, similar to M82 which is dominated by PAH emission in
  the 6-8$\mu$m rest frame, or strong silicate absorption, such as
  Arp\,220, then the f$_{16 \mu m}$/f$_{22 \mu m}$ ratio may vary up
  to a factor of 8 as these features pass in and out of the filters,
  and this ratio can potentially be used as a tracer of the redshift
  of the source. The measured ratios for our sources plotted against
  their minimum possible redshift are indicated by the source ID (see
  Section 3).}

\end{figure*}

\end{document}